\documentclass[showpacs]{article}
\pdfoutput=1

\usepackage{xspace}
\usepackage{textcomp}
\usepackage{makecell}
\usepackage{jcappub}

\newcommand{\sigmasi}{\sigma_\mathrm{S.I.}}

\newcommand{\vesc}{$v_\text{esc} $}
\newcommand{\gevcsqraw}{\text{GeV}/c^2}
\newcommand{\gevcsq}{$\gevcsqraw$\xspace}
\newcommand{\gevpercmcubed}{$\text{GeV}/\text{cm}^3$\xspace}
\newcommand{\kev}{$\text{keV}$\xspace}
\newcommand{\kevee}{$\text{keV}_\text{ee}$\xspace}
\newcommand{\kevnr}{$\text{keV}_\text{nr}$\xspace}
\newcommand{\kever}{$\text{keV}_\text{er}$\xspace}
\newcommand{\ev}{$\text{eV}$\xspace}
\newcommand{\Enr}{E_\text{nr}}
\newcommand{\Eer}{E_\text{er}}
\newcommand{\Eee}{E_\text{ee}}
\newcommand{\cmsqraw}{\text{cm}^2}
\newcommand{\cmsq}{$\cmsqraw$\xspace}
\newcommand{\kmpersec}{$\text{km}/\text{s}$\xspace}
\newcommand{\cross}{cross-section}
\newcommand{\scinot}[1]{$\cdot 10^{#1}$}

\newcommand{\figref}[1]{\autoref{#1}}
\newcommand{\largefigsize}[0]{0.7}
\newcommand{\bigfigsize}[0]{0.8}

\title{Complementarity of direct detection experiments in search of light Dark Matter}
\author[1,2]{J. R. Angevaare,}
\author[1]{G. Bertone,}
\author[1,2]{A. P. Colijn,}
\author[1,2]{M. P. Decowski,}
\author[3]{and B. J. Kavanagh}

\affiliation[1]{GRAPPA, the University of Amsterdam, Science Park, 1098XG Amsterdam, Netherlands}
\affiliation[2]{Nikhef, Science Park, 1098XG Amsterdam, Netherlands}
\affiliation[3]{Instituto de F\'isica de Cantabria (IFCA, UC-CSIC), Avenida de
    Los Castros s/n, 39005 Santander, Spain}
\emailAdd{j.angevaare@nikhef.nl}
\date{\today}

\abstract{
    Dark Matter experiments searching for Weakly interacting massive particles (WIMPs) primarily use nuclear recoils (NRs) in their attempt to detect WIMPs.
    Migdal-induced electronic recoils (ERs) provide additional sensitivity to light Dark Matter with $\mathcal{O}(\text{GeV}/c^2)$ masses.
    In this work, we use Bayesian inference to find the parameter space where future detectors like XENONnT and SuperCDMS SNOLAB will be able to detect WIMP Dark Matter through NRs, Migdal-induced ERs or a combination thereof.
    We identify regions where each detector is best at constraining the Dark Matter mass and spin independent cross-section and infer where two or more detection configurations are complementary to constraining these Dark Matter parameters through a combined analysis.
}

\notoc 
\begin{document}
    \maketitle
    \flushbottom


    \section{Introduction}

    Many Dark Matter direct detection experiments aim to observe Dark Matter (DM) through an excess of nuclear recoils (NRs) caused by Weakly Interacting Massive Particles (WIMPs) scattering off nuclei from a target material~\citep{Drukier:1984vhf, Goodman:1984dc, Drukier:1986tm,Billard:2021uyg}.
    For light Dark Matter, this is not always the most sensitive method of detection.
    For example, the dual-phase liquid xenon experiment XENON1T has reached world-leading sensitivities for a broad range of WIMP-masses using NRs~\citep{Xe1T_one_ton_year} but sensitivity drops quickly for WIMP masses $\lesssim{5}~\gevcsqraw{}$ as the kinetic energy of the WIMP is not sufficient to generate a detectable recoil.
    The lower energy NRs for lighter WIMP-masses typically produce fewer photons and the signal drops below the detection threshold.
    In contrast, cryogenic semiconductor experiments like the Super Cryogenic Dark Matter Search at Sudbury Neutrino Observatory Lab (SuperCDMS)~\citep{superCDMSsnolab} are much better suited for detecting such light DM, due to a combination of a lighter target element, a low energy threshold, and an excellent energy resolution.

    The Migdal effect~\citep{Vergados:2005dpd, Moustakidis:2005gx, Bernabei:2007jz, Ibe_migdal,Dolan:2017xbu,Knapen:2020aky} is a rare, inelastic scattering process that allows the transfer of more energy to the target than with an ordinary NR.
    When an NR causes displacement of the nucleus with respect to the electrons of the atom, the resulting perturbation to the electric field experienced by the electrons may cause ionization or excitation of the atom.
    As such the Migdal manifests itself as an NR causing an electronic recoil (ER).
    While it has not been experimentally confirmed, it offers the possibility for experiments to extend their DM search region to lower WIMP masses~\citep{Xe1T_migdal,PhysRevD.99.082003,CDEX:2019hzn,SuperCDMS:2022kgp,2022arXiv220303993A, Akerib_2019} since NRs that fall below the NR energy threshold of an experiment may result in detectable ERs.

    This paper demonstrates the capability of experiments like XENONnT \citep{XEnT_sens} (the upgrade of XENON1T) and SuperCDMS to reconstruct light Dark Matter, through a combination of NR and Migdal searches.
    Furthermore, we show how the combination of the two experiments would further improve the reconstruction of the DM properties.
    We benchmark the sensitivity of a given detection channel by simulating low mass WIMP signals.
    We then use Bayesian inference to reconstruct the simulated WIMP mass and \cross{}.
    By combining the likelihoods of the two experiments, we study their complementarity.

    References~\citep{Pato_complementarity,Peter:2013aha} have previously demonstrated how experiments employing different target materials such as germanium, xenon and argon could complement each other when using an NR search to reconstruct the Dark Matter mass and \cross{}.
    Additionally, the effect of uncertainties of astrophysical parameters on the reconstruction was investigated (see for example Refs.~\cite{Fox:2010bz,Kavanagh:2013wba}).
    In this work, we will take into account more recent detector characteristics specifically aimed at detecting light Dark Matter through NRs or Migdal analyses.

    In the following section (\autoref{sec:thy}), we review the theory of the NR and Migdal processes.
    The methods section (\autoref{sec:method}) discusses the XENONnT and SuperCDMS detectors, after which the statistical inference framework is introduced.
    In the results section (\autoref{sec:results}) we show the posterior distributions for several benchmarks of interest which we then generalize by exploring the parameter space for WIMP-masses between $0.1-10~\gevcsqraw{}$ and we conclude by summarizing the results (\autoref{sec:conclusion}).


    \section{Theory\label{sec:thy}}

    \subsection{Nuclear recoils}
    The elastic recoil spectrum caused by a WIMP of mass $M_\chi$ scattering off a target nucleus $N(A,Z)$ with mass $M_N$ is described by the differential recoil rate~\citep{Pato_complementarity}:
    \begin{equation}
        \frac{dR}{d\Enr} \left(\Enr\right) =
        \frac{\rho_0}{M_\chi M_N}
        \int\limits_{v_\text{min}}^{v_\text{max}}
        d^3\vec{v}vF\left(\vec{v}+\vec{v}_e\right)
        \frac{d \sigma_{\chi-N}}{d \Enr}(v, \Enr, A) \,
        ,
        \label{eq:nr_spec}
    \end{equation}
    where $\Enr$ is the nuclear recoil energy, $\vec{v}$ is the WIMP velocity in the detector's rest frame for a Dark Matter model with local Dark Matter density $\rho_0$, $\vec{v}_e$ is the Earth's velocity with respect to the galactic rest frame, $F(\vec{v})$ the WIMP velocity distribution in the galactic rest frame and $\sigma_{\chi-N}$ is the WIMP-nucleus \cross{}.
    We will use the same formulation of $\sigma_{\chi-N}$ as in Ref.~\citep{Pato_complementarity}, and only take the spin-independent WIMP-nucleus \cross{} ($\sigmasi$) into account.
    The upper integration limit $v_\text{max}$ is given by the sum of the Dark Matter escape velocity \vesc{} and $\vec{v}_e$.
    The lower integration limit $v_\text{min}$ is the minimum WIMP velocity required to generate an NR of energy $\Enr$.
    The value of $v_\text{min}$ is kinematically constrained and dependent on the target material and recoil energy,
    \begin{equation}
        v_\text{min}\left(\Enr,M_\chi,A\right) =
        \sqrt{\frac{M_N \Enr}{2 \mu_N^2}}
        \label{eq:v_min} \,
        ,
    \end{equation}
    where $\mu_N = \frac{M_\chi M_N}{M_\chi+M_N}$ is the reduced mass and $A$ the atomic mass number of $N(A,Z)$.
    From Eq.~\eqref{eq:nr_spec} we see that for a given recoil rate, a degeneracy exists between $\sigma_{\chi-N}$ and $M_\chi$.
    However, since $v_\text{min}$ also depends on $M_\chi$, this degeneracy may be broken.
    Only when $M_\chi\gg M_N$, Eq.~\eqref{eq:v_min} becomes effectively independent of $M_\chi$, at which point Eq.~\eqref{eq:nr_spec} becomes degenerate for the \cross{} and WIMP-mass.

    In the case of non-directional detectors like XENONnT and SuperCDMS, we can simplify Eq.~\eqref{eq:nr_spec} using the Dark Matter speed distribution $f(v)=4\pi v^2 F(v)$ and ignoring annual modulation effects due to the Earth's orbit around the Sun,
    \begin{equation}
        \frac{dR}{d\Enr} \left(\Enr\right) =
        \frac{\rho_0}{M_\chi M_N}
        \int\limits_{v_\text{min}}^{v_\text{esc}}
        dv \text{ } vf\left(|\vec{v}+\vec{v}_e|\right)
        \frac{d \sigma_{\chi-N}}{d \Enr}(v, \Enr, A) \,
        .
        \label{eq:nr_spec_no_vec}
    \end{equation}
    Earth's velocity relative to the galactic rest frame $\vec{v}_e$ relates to the velocity with respect to the local standard of rest ($\vec{v}_\text{lsr}$), the peculiar velocity ($\vec{v}_\text{pec}$) of the Sun with respect to $\vec{v}_\text{lsr}$ and Earth's velocity ($\vec{v}_\text{Earth-Sun}$) via
    \begin{equation}
        \vec{v}_e =
        \vec{v}_\text{lsr} +
        \vec{v}_\text{pec} +
        \vec{v}_\text{Earth-Sun} \simeq
        \vec{v}_\text{lsr}=
        \vec{v}_0
        \,
        ,
        \label{eq:velocities}
    \end{equation}
    where we have approximated $\vec{v}_e \simeq \vec{v}_\text{lsr}$ which will be referred to as $\vec{v}_0$ throughout this work~\cite{2014JCAP...02..027M}.

    We use a Maxwellian velocity distribution for the Dark Matter velocity distribution $F(v)$, also referred to as the Standard Halo Model \citep{GREEN_2012}.
    For the astrophysical parameters we assume $v_0=233$ \kmpersec, $v_\text{esc}=528$ \kmpersec{} and $\rho_0=0.55$~\gevpercmcubed{}~\citep{Evans_shm}.
    This Dark Matter density $\rho_0$ is different from the 0.3~\gevpercmcubed{} usually assumed for direct detection Dark Matter experiments \citep{lewin1996review, PhysRevLett.116.071301, Xe1T_one_ton_year} which is adopted by convention as its value is directly proportional to the recoil rate as in Eq.~\eqref{eq:nr_spec} and can therefore be easily scaled.
    Ref.~\citep{de_Salas_2021} provides an overview of recent publications on $\rho_0$ where ranges of $0.4-0.6$ and $0.3-0.5$~\gevpercmcubed{} are quoted depending on the type of analysis.
    Using Eqs.~(\ref{eq:nr_spec}-\ref{eq:velocities}), the differential NR rate can be computed for a given target material and a set of astrophysical parameters.

    \subsection{Migdal}

    For lower mass WIMPs, fewer NR energies exceed the energy threshold.
    However, low-energy recoil interactions may be detected through the so-called Migdal effect.
    Although it is usually assumed that the electrons after an NR interaction always accompany the nucleus, it actually takes some time for the electrons to catch up, resulting in ionization and excitation of the recoil atom~\citep{Ibe_migdal}.
    These effects can lead to detectable energy deposits in a detector similar to the energy depositions caused by ERs.
    The differential recoil rates are calculated for several materials assuming isolated atoms in Ref.~\cite{Ibe_migdal}. 
    For semiconductors, the calculation of the Migdal-induced rates  needs to go beyond this isolated atom approximation as was done in Ref.~\citep{Knapen:2020aky}. 
    
    In the isolated atom approximation of Ref.~\cite{Ibe_migdal}, the differential rate for Migdal-induced signals combines the standard NR recoil energy distribution with the electronic band structure of the target atoms.
    The differential Migdal rate is described by the convolution of the NR differential rate with the probability of ionization,
    \begin{equation}
        \frac{dR}{d\Eer} \simeq
        \int d\Enr dv \frac{d^2R}{d\Enr dv} \left(\Enr\right)
        \times
        \sum\limits_{n,l} \frac{d}{d\Eer} P_{q_e}^c
        \left(n, l\to \Eer - E_{n,l}\right)\,,
        \label{eq:migdal_recoil}
    \end{equation}
    where $P_{q_e}^c$ is the probability for an atomic electron with quantum numbers $(n,l)$ and corresponding energy $E_{n,l}$ to be emitted with a kinetic energy of $\Eer - E_{n,l}$.
    The values of $P_{q_e}^c$ are taken from Ref~\citep{Ibe_migdal}.
    
    Ref.~\citep{Knapen:2020aky} includes a derivation of the Migdal-induced rates in semiconductors for WIMP-nucleus scattering.
    Because of the smaller gap for electron excitations, the Migdal rates are found to be higher than for the isolated atom approximation. 
    The differential electronic recoil rate is
    \begin{equation}
        \frac{dR}{d\Eer} \simeq
        \frac{\rho_0}{M_\chi M_N}
        \frac{4\alpha Z^2}{3\pi^2\Eer^4 M_N}\int dk\,k^2\mathrm{Im}\left(\frac{-1}{\epsilon(k, \Eer)}\right)
        \int\limits_{v_\text{min}}^{v_\text{max}}
        d^3\vec{v}vF\left(\vec{v}+\vec{v}_e\right)
        \int d\Enr\,\Enr\frac{d\sigma_{qe}}{d\Enr}
        \,,
        \label{eq:migdal_recoil_semiconductor}
    \end{equation}
    where $\alpha$ is the fine structure constant, $\frac{d\sigma_{qe}}{d\Enr}$ the  quasi-elastic \cross{} from \cite{Knapen:2020aky}, $\mathrm{Im}(-\epsilon^{-1}(k, \Eer))$ the energy loss function with $\epsilon$ the momentum and frequency dependent longitudinal dielectric function, and $k$ is the momentum associated with the electronic excitation.
    
    Using the Migdal effect, the NRs that fall below the energy threshold of experiments may still be indirectly detected as ERs.
    In other words, there is the possibility to detect NRs that are below the threshold through the associated ERs, thereby allowing detectors to be sensitive to smaller WIMP masses that would otherwise be undetectable.


    \section{Methods\label{sec:method}}

    \begin{table*}[t]
        \begin{tabular}{ r | c | c | c | c | c }
            Experiment                      & XENONnT       & \multicolumn{4}{c}{SuperCDMS } \\
            &               & Ge HV     & Si HV     & Ge iZIP       & Si iZIP \\
            \hline\hline
            \multicolumn{6}{l}{\textbf{NR and Migdal (ER)}}\\
            \hline
            Target mass (kg)                & 4\scinot{3}   & 11        &  2.4      & 14            & 1.2 \\
            \hline
            Live time                       & 100\%         & 80\%      &  80\%     & 80\%          & 80\% \\
            \hline
            Run time (yr)                   & 5             & 5         & 5         & 5             & 5 \\
            \hline
            Exposure (kg $\cdot$ {year})        & 20\scinot{3}  & 44        &  9.6      & 56            & 4.8 \\
            \hline
            $k$-parameter for Eq.~\eqref{eq:lindhard}
            & $0.1735$
            & $0.162$
            & $0.161$
            & $0.162$
            & $0.161$ \\
            \hline
            \multicolumn{6}{l}{\textbf{NR}}\\
            \hline\hline
            $E_\text{range}$  ({keVnr})     & [0,~5]        & [0,~5]    & [0,~5]    & [0,~5]        & [0,~5] \\
            \hline
            Cut- and detection-eff.   & 0.83          & $0.85\cdot0.85$& $0.85\cdot0.85$ & $0.85\cdot 0.75$ &$0.85\cdot 0.75$\\
            \hline
            Energy resolution               & Eq.~\eqref{eq:det_res_Xe_nr}
            & Eq.~\eqref{eq:sigma_ph_nr}
            & Eq.~\eqref{eq:sigma_ph_nr}
            & Eq.~\eqref{eq:sigma_q_nr}
            & Eq.~\eqref{eq:sigma_q_nr} \\
            for $\sigma_\text{ph, nr}$ (HV) / $\sigma_{Q,\,\text{nr}}$ (iZIP) &
            & $10~$\ev{}
            & $5~$\ev{}
            & $100~$\ev{}
            & $110~$\ev{}\\
            \hline
            \makecell[r]{BG. $\left(\frac{\text{counts}}{{\text{kg} \cdot \text{keV} \cdot \text{year}}}\right)$}
            & 2.2\scinot{-6}& 27
            &  300
            &3.3\scinot{-3}
            & 2.9\scinot{-3} \\
            \hline
            $E_\text{thr}$~(\kevnr)          & 1.6           & 0.040     & 0.078     & 0.272         & 0.166\\
            \hline\hline
            \multicolumn{6}{l}{\textbf{Migdal (ER)}}\\
            \hline
            $E_\text{range}$~(\kevee)        & [0,~5]        & [0,~0.5]    & [0,~0.5]    & [0,~0.5]        & [0,~0.5] \\
            \hline
            Cut- and detection-eff.   & 0.82
            & $0.5\cdot0.85$
            & $0.675\cdot0.85$
            & $0.5\cdot 0.75$
            & $0.675\cdot 0.75$\\
            \hline
            Energy resolution               & Eq.~\eqref{eq:det_res_Xe}
            & 0.4~$\text{eV}_\text{ee}$
            & 0.15~$\text{eV}_\text{ee}$
            & 19~$\text{eV}_\text{ee}$
            & 7~$\text{eV}_\text{ee}$ \\
            \hline
            \makecell[r]{BG. $\left(\frac{\text{counts}}{{\text{kg} \cdot \text{keV} \cdot \text{year}}}\right)$}
            & 12.3\scinot{-3}
            & 27
            & 300
            & 22
            & 370\\
            \hline
    $E_\text{thr}$  (keVee)        & 1.0           & 0.004      & 0.003     & 0.14          & 0.05\\
            \hline
        \end{tabular}
        \caption{
            The assumed detector characteristics of XENONnT and SuperCDMS.
            SuperCDMS consists of various detector target materials (Si, Ge) and designs (HV, iZIP).
            The first set of detector parameters (top part of the table) are independent of the type of analysis (NR or Migdal).
            For the NR and Migdal searches, the respective values are listed separately in the middle and bottom of the table.
            \label{tab:det_params}}
        \normalsize
    \end{table*}

    We consider two experiments: XENONnT and SuperCDMS.
    These detectors are both sensitive to $\mathcal{O}\left(\gevcsqraw{}\right)$ mass WIMPs, but with significant differences: SuperCDMS has a high quantum yield with a relatively modest target mass, while XENONnT combines a lower light and charge yield with a multi-tonne target mass.

    In the remainder of this section, we describe the methods we use for modeling the detectors, calculating the signal spectra, and inferring projected constraints on the DM parameters.
    The detector characteristics which are used are summarised in \autoref{tab:det_params}.
    Example NR and Migdal spectra for the experiments are shown in \autoref{fig:recoil_spectra}.
    We use \texttt{pymultinest} to sample from the posterior distribution of the spin-independent WIMP-nucleon \cross{} and WIMP mass ($\sigmasi,\,M_\chi$), assuming the benchmark points and priors given in \autoref{tab:benchmarks}.
    The results of these benchmark points are further generalized in the Results section (\autoref{sec:results}).

    For both experiments we assume a five-year run time which the experiments aim to acquire on similar timescales~\citep{superCDMSsnolab, XEnT_sens}.
    The product of a combined cut- and detection- efficiency, run time, live time and target mass yields the effective exposure $\epsilon_\text{eff}$.

    Below, we describe the detector characteristics which are used for the recoil rate calculations, summarized in \autoref{tab:det_params}.
    In the following sections, we use the Lindhard theory~\citep{lindhard1963integral} to convert between NR energies ($\Enr$) and electronic equivalent energies ($\Eee$) as explained in Appendix~\ref{ap:lindhard}.
    For both the NR and Migdal search, we require the cut- and detection-efficiency, energy resolution, background rate, and energy thresholds for the calculation of the spectra.
    As the Migdal effect manifests itself as an ER signal, some parameters are different from the NR search, such as the expected background in case the detector has the ability to distinguish NRs and ERs.
    Other parameters like target mass and exposure are independent of the type of search.
    We conclude this section with a description of the Bayesian framework we use for the analysis.

    \subsection{XENONnT}

    XENONnT is the upgrade of XENON1T with a larger target mass and lower background expectation~\citep{XEnT_sens}.
    For the NR and Migdal detection channels, we assume a 4~tonne active target mass and continuous data taking (live time of 100\%), yielding a total of 20~tonne~year exposure.

    XENONnT measures both prompt scintillation light (S1) and ionization signals (S2).
    Since NRs with the same energy cause relatively smaller ionization signals, XENONnT is able to distinguish between ERs and NRs.
    Most of the background events in XENONnT are from radioactive contaminants like radon and krypton causing ERs within the active target volume.
    The background rate for the NR search can therefore be reduced because of the ER/NR discrimination.
    We assume a background rate of $2.2\cdot10^{-3}$~(12.6)~$\text{keV}^{-1}\text{t}^{-1}\text{yr}^{-1}$ for the NR (Migdal) search~\citep{XEnT_sens}.
    We will first discuss the parameters relevant for the Migdal search followed by those for the NR search.

    For the Migdal search, the detector ER energy resolution ($\sigma$ in \kever) is assumed to be the same as for XENON1T \cite{Xe1T_lower} which is given by the empirical formula:
    \begin{equation}
        \sigma_\text{er}(\Eer) = 0.31\,\text{keV}_\text{er}\sqrt{\frac{\Eer}{\text{keV}_\text{er}}} + 0.0037\, \Eer
        \label{eq:det_res_Xe}
        \,
        .
    \end{equation}
    The ER detection energy threshold relevant for the Migdal search ($E_\text{thr,\,er}$) is assumed to equal 1.0~\kever{}~\citep{Xe1T_lower}.
    This energy threshold is dictated by the requirement of reconstructing the S1 of an interaction~\citep{Xe1T_migdal}.
    While lower thresholds are achieved in S2-only analyses, these can only lead to exclusion of Dark Matter models as not all backgrounds can be adequately modelled~\citep{aprile2019light}. 
    Therefore, these lower thresholds are not used here.

    The Migdal recoil energies are limited to the interval of [0,~5]~\kever.
    While Ref.~\citep{Ibe_migdal} assumes target materials to consist of isolated atoms, XENONnT uses liquid xenon as the target material. 
    To account for this difference and in order to be conservative, the contribution to the differential recoil rate from the $5,1$ shell is neglected. 
    We do take the $5,0$ shell into account which contributes $\lesssim2\%$ to the total rate for the masses considered in this work.
    Furthermore, the innermost electrons are considered too tightly bound to the nucleus to contribute significantly~\citep{Ibe_migdal, Xe1T_migdal, PhysRevD.99.082003}.
    Finally, we assume a combined detection and cut efficiency of $83\%$~($82\%$) for NR (Migdal)~\citep{XEnT_sens}.

    For the NR search, we use the Lindhard factor $L$ (explained in \autoref{ap:lindhard}) in Eq.~\eqref{eq:ee_to_nr} to convert $\Enr$ to $E_\text{ee}$ and treat the energy resolution (Eq.~\eqref{eq:det_res_Xe}) as the uncertainty on the value of the detected energy:
    \begin{equation}
        \sigma_\text{nr}(\Enr)
        =
        \frac{\text{d} \Enr}{\text{d} \Eer}\sigma_\text{er}(E_\text{ee})
        =
        \frac{\text{d} \Enr}{\text{d} \Eer}\sigma_\text{er}\left(L(\Enr)\cdot \Enr\right)\,,
        \label{eq:det_res_Xe_nr}
    \end{equation}
    to obtain the NR energy resolution $\sigma_\text{nr}$.
    A value of $k=0.1735$ \citep{akerib2016low} is used for XENONnT in Eq.~\eqref{eq:lindhard}.
    We assume an analysis optimized for low energy events.
    We set an energy threshold $E_\text{thr,\, nr}$ of $1.6$~\kevnr{}, which has been achieved in XENON1T with the dedicated low energy NR search for coherent elastic scattering of solar neutrinos~\citep{cevns_1t}.
    The energy range of interest is set to [0,~5]~\kevnr.

    \subsection{SuperCDMS}

    The SuperCDMS experiment \cite{superCDMSsnolab} has two detector designs each using germanium and silicon as target material.
    The so-called HV detector only utilizes phonon sensors, whereas the iZIP detector uses both phonon and ionization sensors, thereby allowing ER/NR discrimination.
    Since the HV detectors are not able to distinguish between ER and NR, most of the detector parameters are the same for the Migdal (ER) and NR search.
    For the iZIP detectors some detector parameters differ for the two types of searches because of the ER/NR discrimination.

    The HV detectors have better phonon energy resolution compared to the iZIP detectors, which results in a better sensitivity for WIMP masses $\lesssim5~\gevcsqraw{}$ as lower WIMP masses cause lower recoil energies.
    The iZIP detectors have better sensitivity for higher masses.
    We model each of the target materials for each of the detector designs, yielding four different configurations.
    The detector parameters are listed in \autoref{tab:det_params}.

    The background in each detector is directly obtained from Table V. in Ref.~\citep{superCDMSsnolab}.
    The backgrounds of the HV detector (NR and Migdal search) are given by the ER backgrounds dominated by $^{3}$H and $^{32}$Si decays.
    The iZIP detector background for Midgal is also given by the ER background whereas the NR search background, which is mostly due to coherent neutrinos, is significantly lower due to the NR/ER discrimination.

    The energy-scales, -resolution and -thresholds for the four detector configurations for both NR and Migdal are summarized in Appendix~\ref{ap:e_supercdms}.
    Their respective values are listed in \autoref{tab:det_params}.
    For the NR search, we use a [0,~5]~\kevnr energy range.
    As the electronic recoil energies for the Migdal search are typically at low energy, we focus on the energy range of [0,~0.5]~\kever.

    \subsection{Recoil rates}

    \begin{figure*}[t]
        \centering
        \includegraphics[width=\bigfigsize\textwidth]{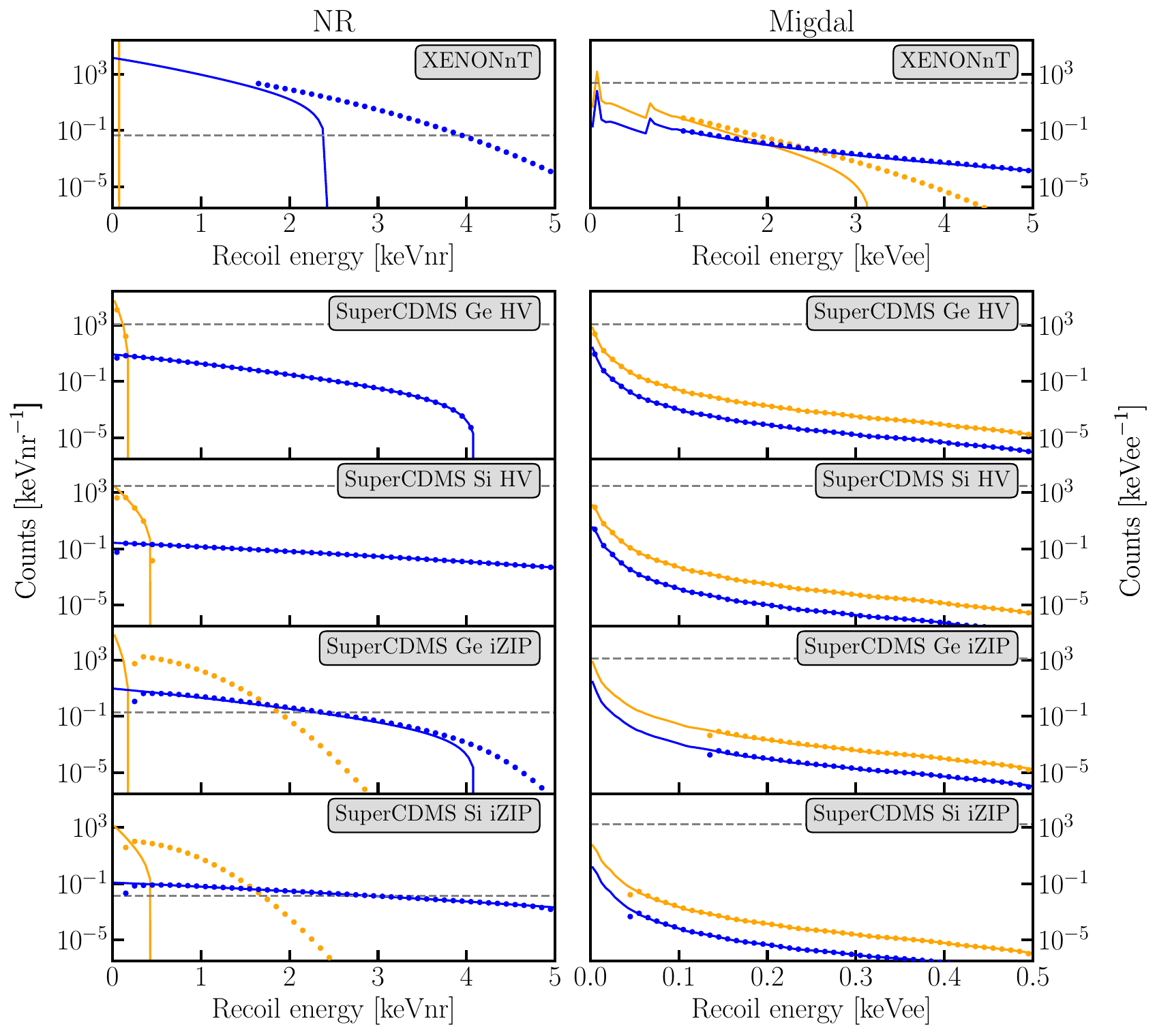}
        \caption{
            Recoil spectra for WIMP DM with $M_\chi=5$~\gevcsq{} and $\sigmasi=10^{-45}$~\cmsq{} (blue) and $M_\chi=1$~\gevcsq{} and $\sigmasi=10^{-42}$~\cmsq{} (orange) for the exposures listed in \autoref{tab:det_params}.
            The differential recoil rate (solid line) results in the detectable spectrum (dots) when the detector energy threshold and detector resolution are taking into account, and the spectrum is binned in 50 energy bins.
            The background rates for the given exposures are shown separately (dashed gray lines).
            The left column shows the NR spectra and the right column the ER spectra as a result of the Migdal effect.
            For all NR searches, the energy range is restricted to $[0,~5]$~\kevnr{}, while for Migdal the SuperCDMS searches use a smaller energy interval of $[0,~0.5]$~\kever{} compared to XENONnT ($[0,~5]$~\kever{}).
            In the XENONnT-NR panel, the recoil rate for $M_\chi=1$~\gevcsq{} falls off exponentially well below the energy threshold of 1.6~\kevnr{} and the detectable spectrum is $\sim0$~counts \kevnr{}$^{-1}$.
            For example for the XENONnT detector, especially with $M_\chi=1$~\gevcsq{}, the top panels show why the Migdal effect can help experiments extend their search region, since even though the spectrum drops steeply below the NR energy threshold, the Migdal spectrum extends sufficiently beyond the detector energy threshold of 1.0~\kevee{} to higher ER energies.
            \label{fig:recoil_spectra}
        }
    \end{figure*}

    In order to evaluate the recoil spectra, we evaluate  Eq.~\eqref{eq:nr_spec} or Eq.~\eqref{eq:migdal_recoil} using the \texttt{wimprates}-framework~\citep{wimprates} and Eq.~\eqref{eq:migdal_recoil_semiconductor} using the \texttt{darkelf}-framework~\citep{Knapen:2020aky,Knapen_2022}.
    For evaluating the energy loss function in Eq.~\eqref{eq:migdal_recoil_semiconductor}, we use the GWAP method for $\Eer<60~\ev{}$ and Lindhard method for larger energies as no data for the GPAW~\citep{Knapen_2022} method is available at energies $\Eer\gtrsim75~\ev{}$ and the methods agree well for recoils above $60~\ev$.
    To calculate the recoil rates, we assume the astrophysical parameters as per the Standard Halo Model.
    We will limit ourselves to WIMPs that couple to the target nucleus through spin-independent interactions.

    We add a flat background spectrum to the NR or Migdal recoil spectrum prior to convolving the spectrum with the detector resolution $\sigma$, resulting in the detectable energy spectrum
    \begin{equation}
        \frac{d\tilde{R}}{dE_R} =
        \int dE' \frac{dR}{dE_R}(E')
        \frac{e^{-\frac{(E-E')^2}{2\sigma^2(E')}}}{\sqrt{2\pi} \sigma(E')}
        \,.
        \label{eq:res_smearing}
    \end{equation}
    The number of expected events $N_i$ in a given energy bin is obtained by integrating Eq.~\eqref{eq:res_smearing} times the effective exposure ($\epsilon_\text{eff}$) between the bin edges $E_\text{min}^i,~E_\text{max}^i$,
    \begin{equation}
        N_i = \int_{E_\text{min}^i}^{E_\text{max}^i} d E_R \epsilon_\text{eff} \frac{d\tilde{R}}{d E_R}
        \,.
        \label{eq:Ni}
    \end{equation}

    \figref{fig:recoil_spectra} shows the spectra obtained for NR and Migdal before- and after- including detector effects as well as the background rates for each detector.
    We approximate the spectrum by a 50-bin spectrum which allows for reasonably fast computation of spectra.

    We model the Migdal spectra and NR spectra independent from each other.
    In a real detector when DM would be observed through the Migdal effect, the direct NRs may also be observed.
    This is especially relevant for detectors where there is no NR/ER discrimination as the Migdal and NR contribution could not be disentangled.
    Since we want to investigate the ability of detectors to detect DM through either Migdal or NR, we take their resultant spectra separately into account as if only one or the other would be observed.

    \subsection{Statistical inference \label{sec:stat_infer}}

    We follow a Bayesian approach~\cite{Bayes:1764vd} to extract the parameters of interest ($M_\chi$ and $\sigmasi$) similar to the method described in Ref.~\citep{Pato_complementarity}.
    The total likelihood $\mathcal{L}$ is the product of the likelihood for each detector which is given by the product of the Poisson probability of each of the energy bins
    \begin{equation}
        \label{eq:bin_likelihood}
        \mathcal{L}\left(\Theta\right) =
        \prod^\text{detectors}_j
        \left(
        \prod_{i}^\text{bins}
        \frac{\hat{N}_{ij}(\Theta)^{N_i}}{N_i!}e^{-\hat{N}_{ij}(\Theta)}
        \right)
        \,
        ,
    \end{equation}
    where $N_i$ is the number of counts in each energy bin~($i$) and $\hat{N}_{ij}(\Theta)$ is the expected counts for a given detector~($j$) at the set of parameters $\Theta$, where $\Theta$ contains the DM parameters of interest,
    \begin{equation}
        \label{eq:parameters_theta}
        \Theta = \{ M_\chi, \sigmasi\}
        \,.
    \end{equation}

    To infer the posterior distribution, the likelihood $\mathcal{L}(\Theta)$ is multiplied by the prior $p(\Theta)$ for given parameters $\Theta$.
    We choose a flat prior in log-space for the mass and \cross{} as their true value is unknown and the aim is to reconstruct these parameters.
    Given the very steep rise in sensitivities for SuperCDMS and XENONnT in the mass range considered here, a large prior range was chosen for the masses of interest.
    Each of the prior ranges was set around the central value for the three benchmark points of interest, as in \autoref{tab:benchmarks}.

    The likelihood for SuperCDMS at $\Theta$ is given by the product of the likelihood of the Ge~HV, Si~HV, Ge~iZIP and Si~iZIP detectors.
    When combining the results of XENONnT and SuperCDMS, all five detectors are taken into account in the product over the detectors in Eq.~\eqref{eq:bin_likelihood}.

    To sample the posterior distribution several sampling methods are implemented in Ref.~\citep{dddm} such as \texttt{emcee}~\citep{emcee}, \texttt{nestle}~\citep{nestle} and \texttt{pymultinest}~\citep{pymultinest}.
    Since the results are independent of the sampling method and \texttt{pymultinest} proved the fastest, it is used here.
    The \texttt{pymultinest}-package is a pythonic interface to the \texttt{multinest} algorithm~\citep{multinest_a, multinest_b}.

    Using the \texttt{pymultinest} sampler, 1000 ``live points" are generated that populate the prior volume.
    The live points iteratively probe the prior volume to obtain the posterior, see Ref.~\citep{multinest_b}.
    A tolerance of 0.5 is used as a stopping criterion.
    The samples are weighted to represent the posterior distribution density.

    \begin{table*}[t!]
        \begin{tabular}{ c | c | c | c }
        {$M_\chi$~$(\gevcsqraw{})$} &
            {$\sigmasi$~(\cmsq{})} &
            {prior-range~ $\log_{10}\left(M_\chi/\left(\gevcsqraw{}\right)\right)$} &
            {prior-range~$\log_{10}\left(\sigmasi/\cmsqraw{}\right)$} \\ \hline
        $5$                & $10^{-45}$  &  $\log_{10}(5)-2.5\text{,}\log_{10}(5)+3.5$     & $-52\text{,}-40 $\\
        $3$                & $10^{-41}$  &  $\log_{10}(3)-2.5\text{,}\log_{10}(3)+3.5$     & $-48\text{,}-36 $\\
        $0.5$              & $10^{-38}$  &  $\log_{10}(0.5)-2.5\text{,}\log_{10}(0.5)+3.5$ & $-45\text{,}-33 $\\
        \end{tabular}
        \caption{Benchmark points and corresponding prior ranges.
        For both the WIMP mass \cross{s} a flat prior is assumed in log-space.
        As the relevant \cross{}s greatly differ for the three WIMP masses, the prior ranges are scaled accordingly.
        \label{tab:benchmarks}}
    \end{table*}


    \section{Results\label{sec:results} and discussion}

    \begin{figure*}[t!]
        \centering
        \includegraphics[width=\largefigsize\textwidth]{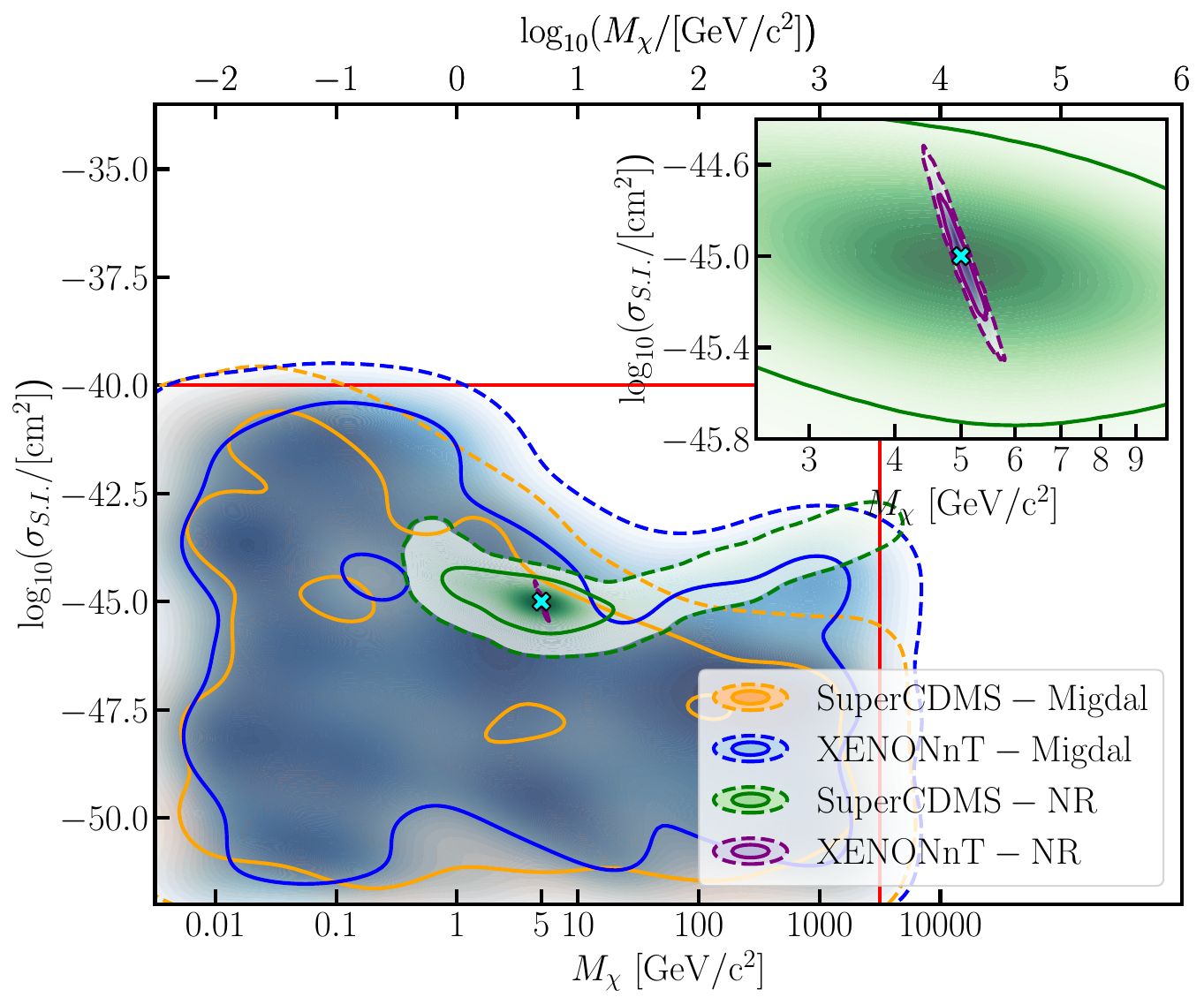}
        \caption{
            \label{fig:nr_vs_migdal_5gev}
            Posterior distribution densities reconstructed for a WIMP with $M_\chi=5$~\gevcsq{} and $\sigmasi=10^{-45}$~\cmsq{} in the four detector configurations.
            The 68\% and 95\% CIs are illustrated with the solid and dashed lines, respectively.
            Whereas the NR searches are able to reconstruct the set benchmark (cyan), the Migdal searches are not.
            The inset shows the posterior distribution densities XENONnT-NR  and SuperCDMS-NR, where the 68\% CI for the former is much smaller than that of the latter.
            The XENONnT-Migdal and SuperCDMS-Migdal reconstructed posteriors fill the prior volume (indicated by the red box), consistent with no signal.
        }
    \end{figure*}

    For a given set of Dark Matter parameters $\Theta$, a benchmark recoil spectrum is calculated for each of the detectors.
    We obtain the posterior distribution density using \texttt{pymultinest} to investigate how a binned Poisson likelihood analysis would be able to reconstruct the set DM parameters.
    This section compares the ability of SuperCDMS and XENONnT to correctly reconstruct $\Theta$ using either an NR or Migdal search.

    SuperCDMS and XENONnT have different characteristics (\autoref{tab:det_params}) and their ability to reconstruct the benchmark value depends strongly on the assumed DM parameters.
    We give results for the three benchmark points in \autoref{tab:benchmarks} which lie close to the detection threshold of XENONnT.
    Next, we generalize this for other masses and cross-sections to find the complementarity of the four detector configurations.

    \subsection{5 \gevcsq{}}

    We first simulate a benchmark Dark Matter model for WIMPs with $M_\chi=5$~\gevcsq{} and $\sigmasi=10^{-45}$~\cmsq{}.
    \figref{fig:nr_vs_migdal_5gev} shows the inferred posterior distribution for these Dark Matter parameters, which XENONnT NR-search (XENONnT-NR) reconstructs since the benchmark value is in the center of the posterior distribution density.
    Also, the SuperCDMS NR-search (SuperCDMS-NR) gives the Dark Matter parameters albeit with a larger 68\% credibility interval (CI), while at large $M_\chi$ the 95\% CI contour lines do not close due to a mass-\cross{} degeneracy as mentioned in the Theory section (\autoref{sec:thy}).
    The difference between XENONnT-NR and SuperCDMS-NR can be understood from  \figref{fig:recoil_spectra}: the number of expected events for XENONnT-NR for $M_\chi=5~\gevcsqraw{}$ is higher while the background is relatively lower than for SuperCDMS-NR, leading to a tighter 68\% CI for XENONnT-NR.

    The XENONnT Migdal-search (XENONnT-Migdal) and SuperCDMS Migdal-search (Super-CDMS-Migdal) are not able to reconstruct the benchmark point.
    For these detector configurations, the prior volume is filled where the signal would be consistent with no signal, since the expected recoil rates in \figref{fig:recoil_spectra} are relatively low and backgrounds generally higher compared to the NR searches (\autoref{tab:det_params}).
    When the \cross{} and WIMP mass are both higher, a sizable Migdal signal is expected.
    Therefore, the prior volume in the upper right corner of \figref{fig:nr_vs_migdal_5gev} is not filled by the posterior distributions of XENONnT-Migdal and SuperCDMS-Migdal.

    We quantify how well the benchmark is reconstructed by calculating the fraction of the prior volume filled by the posterior volume in log-space of the enclosed 68\% CI:
    \begin{equation}
        \phi =
        \frac{
            \log_{10}
            \left(
            \dfrac
            {M_\chi^\text{enc. 68\%}}
            {\gevcsqraw{}}
            \right)
            \cdot
            \log_{10}
            \left(
            \dfrac
            {\sigmasi^\text{enc. 68\%}}
            {\cmsqraw}
            \right)
        }
        {\text{prior-volume}}
        \label{eq:eta}
        \,,
    \end{equation}
    which is the surface enclosed by the solid lines in \figref{fig:nr_vs_migdal_5gev} divided by the surface within the red box.
    The 68\% CI is obtained using a bi-variate Gaussian kernel density estimator based on code from Ref.~\citep{seaborn}.
    Values of $\phi\sim\mathcal{O}(0.1-1)$ indicate low power to reconstruct a benchmark model since the posterior volume is of similar size as the prior volume, the lower $\phi$, the better the benchmark is reconstructed as the parameters are better constrained.

    Evaluating $\phi$ for the results in \autoref{fig:nr_vs_migdal_5gev} yields $\phi_\textrm{XENONnT-NR}=6.1\times10^{-5}$ while $\phi_\textrm{SuperCDMS-NR}=8.1\times10^{-3}$, showing that the XENONnT-NR search yields $\mathcal{O}(10^2)$ times tighter constraints on the reconstructed parameters.
    For the Migdal searches $\phi$ is large ($\phi_\textrm{XENONnT-Migdal}=3.9\times10^{-1}$) and ($\phi_\textrm{SuperCDMS-Migdal}=3.5\times10^{-1}$).
    As the 95\% CI do not close before the prior boundaries, these numbers only indicate that neither XENONnT-Migdal nor SuperCDMS-Migdal is able to reconstruct the DM parameters.

    \subsection{3 \gevcsq{}}

    \begin{figure*}[t!]
        \centering
        \includegraphics[width=\largefigsize\textwidth]{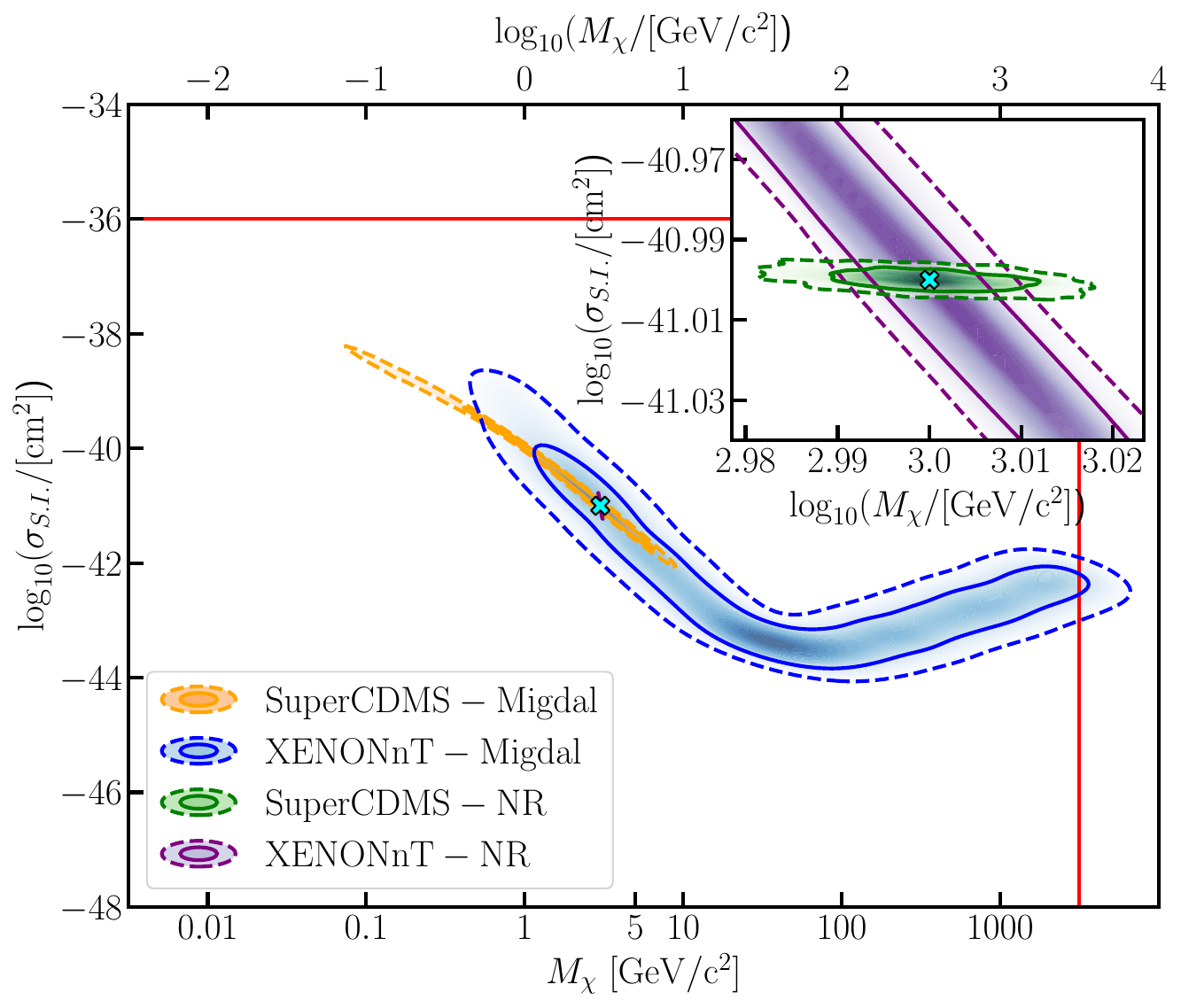}
        \caption{
            \label{fig:nr_vs_migdal_3gev}
            Posterior distributions reconstructed for a WIMP with $M_\chi=3$~\gevcsq{} and $\sigmasi=10^{-41}$~\cmsq{} in the four detector configurations.
            SuperCDMS-NR and XENONnT-NR both reconstruct the benchmark point (cyan) even though the shapes of the posterior differ.
            Furthermore, the SuperCDMS-Migdal is also able to constrain the DM parameters with larger 68\% and 95\% CIs.
            The posterior for XENONnT-Migdal has non-closing contour lines as it extends to the boundary of the prior range as in \autoref{tab:benchmarks}.
        }
    \end{figure*}

    We simulate a WIMP of $M_\chi=3$~\gevcsq{} and $\sigmasi=10^{-41}$~\cmsq{} near the detection threshold of XENONnT.
    At this mass and \cross{}, XENONnT-NR and SuperCDMS-NR both reconstruct a tight posterior distribution as in \figref{fig:nr_vs_migdal_3gev}. 
    As this \cross{} is higher than what was considered for 5~\gevcsq{}, SuperCDMS-Migdal and XENONnT-Migdal are also able to reconstruct a broad posterior distribution which, for XENONnT-Migdal, has non-closing contour lines due to the mass-\cross{} degeneracy also observed for SuperCDMS-NR in \figref{fig:nr_vs_migdal_5gev}.

\begin{figure}[t]
        \centering
        \includegraphics[width=0.525\textwidth]{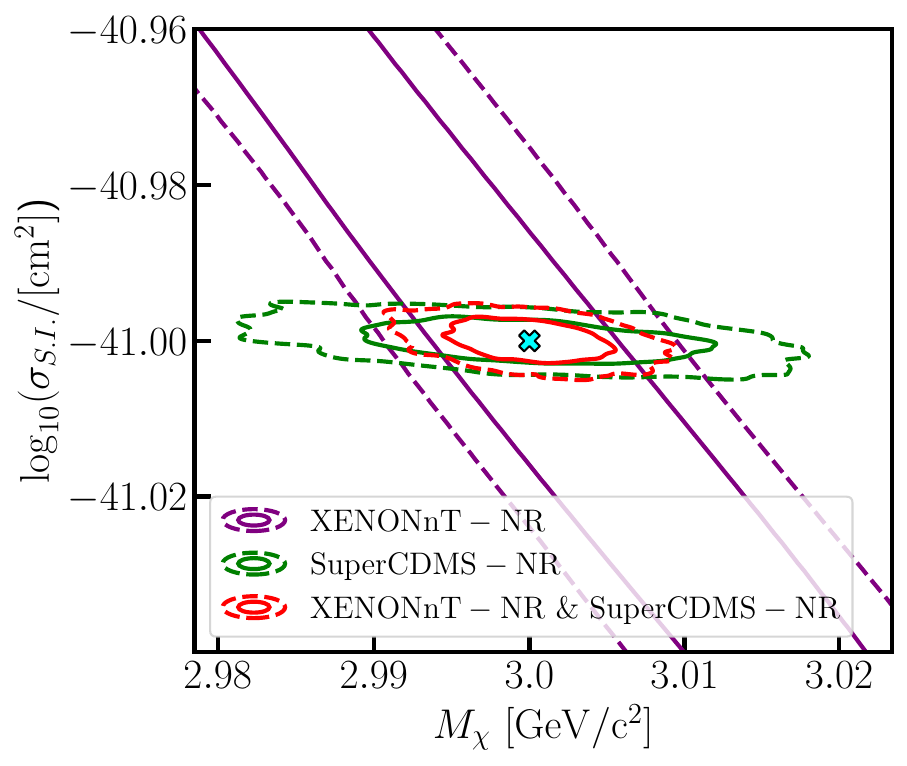}
        \caption{
            \label{fig:nr_vs_migdal_3gev_compare}
            Overlaid posterior distributions reconstructed for a WIMP with $M_\chi=3$~\gevcsq{} and $\sigmasi=10^{-41}$~\cmsq{} for SuperCDMS-NR (green), XENONnT-NR (purple) and the combined result for SuperCDMS-NR and XENONnT-NR (red).
            The 68\% CI (solid) and 95 \% CI (dashed) contour lines are shown.
            The two experiment are complementary to each other since a combination of the two experiments yields a substantially tighter 68\% CI as explained in the text.
        }
    \end{figure}

    We study the complementarity of XENONnT-NR and SuperCDMS-NR in \figref{fig:nr_vs_migdal_3gev_compare}.
    Whereas the reconstructed 68\% CI for XENONnT-NR has a relatively large spread in $\sigmasi$, SuperCDMS-NR has a large spread in $M_\chi$.
    The likelihood of XENONnT-NR changes rapidly as function of $M_\chi$ since the drop in the recoil spectrum occurs close to the energy threshold for these WIMP masses.
    As a result, the likelihood constrains $M_\chi$ around this mass relatively well.
    In contrast, the uncertainty of SuperCDMS-NR is mostly in $M_\chi$ since a shift in the spectral shape as function of $M_\chi$ has a relatively smaller effect for SuperCDMS-NR on the number of events above threshold.
    Since $\sigmasi$ is proportional to the number of events observed it is therefore relatively well constrained for SuperCDMS-NR.

    When the likelihoods of the NR searches are combined, the 68\% CI is reduced.
    Quantitatively, one can see this from
    $\phi_\textrm{XENONnT-NR}=2.8\times10^{-6}$ and $\phi_\textrm{SuperCDMS-NR}=1.1\times10^{-7}$ while the combination of the two gives $\phi_\textrm{XENONnT-NR+SuperCDMS-NR}=5.1\times10^{-8}$.
    This corresponds to a reduction of $\phi$  by a factor of 54 (2.1)   when the likelihoods of these detector configurations are combined, compared to XENONnT-NR (SuperCDMS-NR) alone.
    Both Migdal searches also constrain the posterior distribution, $\phi_\textrm{SuperCDMS-Migdal}=8.8\times10^{-4}$ and $\phi_\textrm{XENONnT-Migdal}=2.3\times10^{-2}$.
    However, since the 68\% CI of SuperCDMS-Migdal and XENONnT-Migdal fully enclose the 68\% CI of the XENONnT-NR search, their combination with the NR searches does not result in a lower value of $\phi$.

    \subsection{0.5 \gevcsq{}}

    \begin{figure*}[t]
        \centering
        \includegraphics[width=\largefigsize\textwidth]{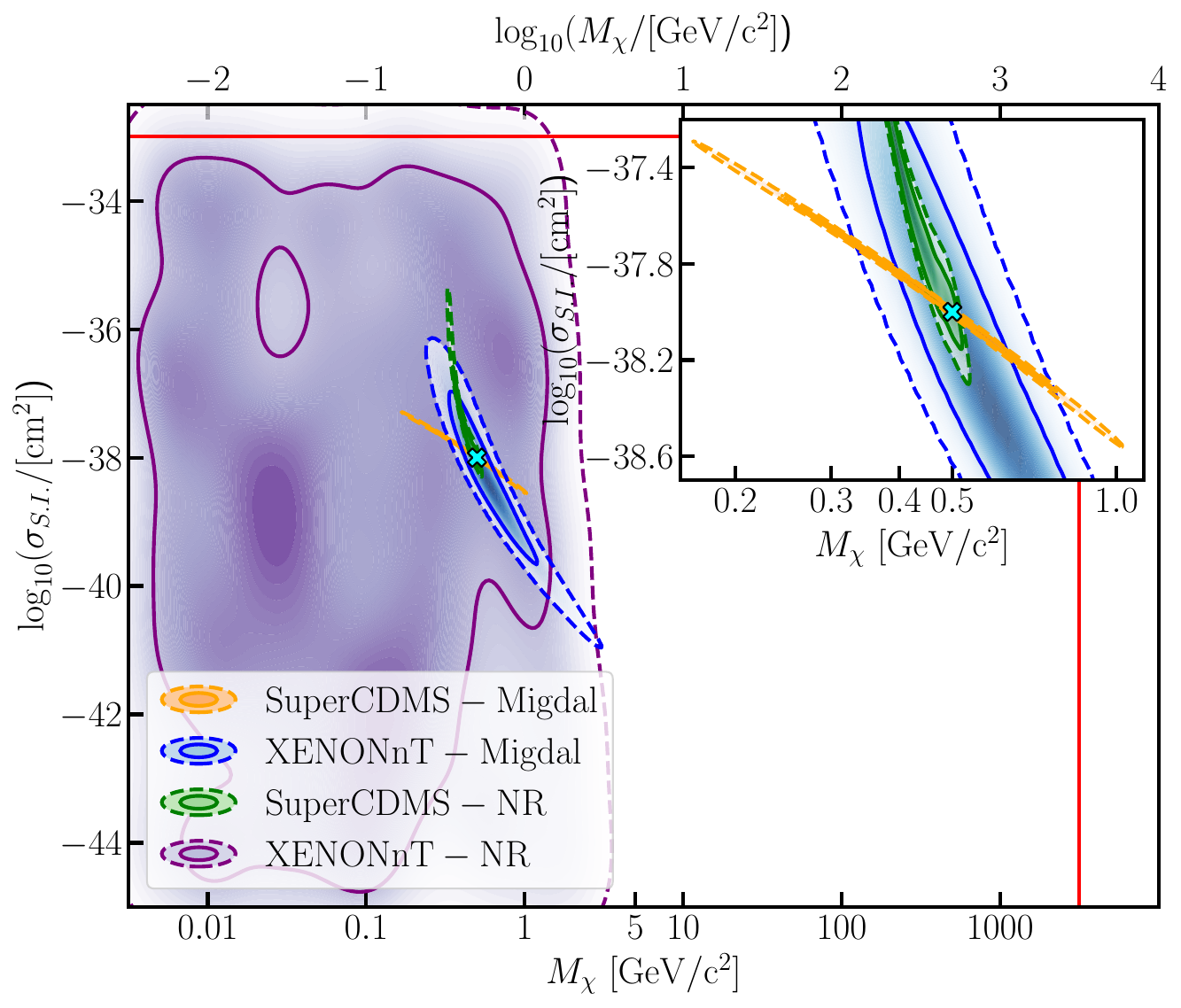}
        \caption{
            \label{fig:nr_vs_migdal_05gev}
            The posterior distributions reconstructed for a WIMP with $M_\chi=0.5$~\gevcsq{} and $\sigmasi=10^{-38}$~\cmsq{}.
            SuperCDMS-NR and SuperCDMS-Migdal reconstruct the benchmark point (cyan) as the 68\% CI (solid) and 95 \% CI (dashed) center around the set benchmark.
            Whereas XENONnT-NR does not reconstruct the benchmark, the Migdal search does.
            Due to the few detected recoils and relatively large background for XENONnT-Migdal, the credibility interval is significantly larger than for SuperCDMS-NR or SuperCDMS-Migdal.
        }
    \end{figure*}

    When considering a lower mass WIMP of $M_\chi=0.5~\gevcsqraw{}$ and $\sigmasi=10^{-38}~\cmsqraw{}$ the situations changes.
    The spectra in \figref{fig:recoil_spectra} are shifted to lower energies and for XENONnT-NR, the spectrum (before taking the detector effects into account) drops steeply below the energy threshold, leading to close to no events in the detector.
    At this \cross{}, the recoil rate for XENONnT-Migdal becomes sufficient to constrain the DM parameters.
    \figref{fig:nr_vs_migdal_05gev} shows the posterior distributions for the four detector configurations.
    
    The SuperCDMS-Migdal search is able to reconstruct these DM parameters best, resulting in $\phi_\textrm{SuperCDMS-Migdal}=6.0\times10^{-5}$.
    The NR search of SuperCDMS also constrains the DM parameters, achieving $\phi_\textrm{SuperCDMS-NR}=2.2\times10^{-4}$.
    The XENONnT-NR search becomes insensitive as fewer signals are above the energy threshold ($\phi_\textrm{XENONnT-NR}=2.3\times10^{-1}$), the posterior distribution function fills the prior volume up to $\sim3$~\gevcsq{}, where NRs are starting to be just above the detection energy threshold.
    In contrast, for such a \cross{} and mass, the XENONnT-Migdal search is able to constrain the posterior distribution ($\phi_\textrm{XENONnT-Migdal}=2.3\times10^{-3}$).
    With the considered $M_\chi$ being close to the energy threshold of SuperCDMS-NR, the 68\% CI of SuperCDMS-NR extends to lower masses and higher \cross{}s with respect to the benchmark point since a higher mass would result in many more events.
    In contrast, the 68\% CI of XENONnT-Migdal is quite broad due to the limited number of events at this \cross{} and mass, while being less affected by the energy threshold.
    Since the 68\% CI of SuperCDMS-NR and XENONnT-Migdal cover different portions of the prior volume the combination of the two has a much lower ($\phi_\textrm{SuperCDMS-NR+XENONnT-Migdal}=3.4\times10^{-5}$), which is a factor of 6 lower than for SuperCDMS-NR and a factor of 69 compared to XENONnT-Migdal.
    Even better results are achieved with the combination of SuperCDMS-NR and SuperCDMS-Migdal, where $\phi_\textrm{SuperCDMS-NR+SuperCDMS-Migdal}=8.1\times10^{-8}$, which corresponds to a reduction of $7\times10^{2}$ for SuperCDMS-Migdal and $3\times10^{3}$ for SuperCDMS-NR.

    \subsection{Masses between 0.1-10~\gevcsq{}}

    \begin{figure*}[t!]
        \centering
        \includegraphics[width=\textwidth]{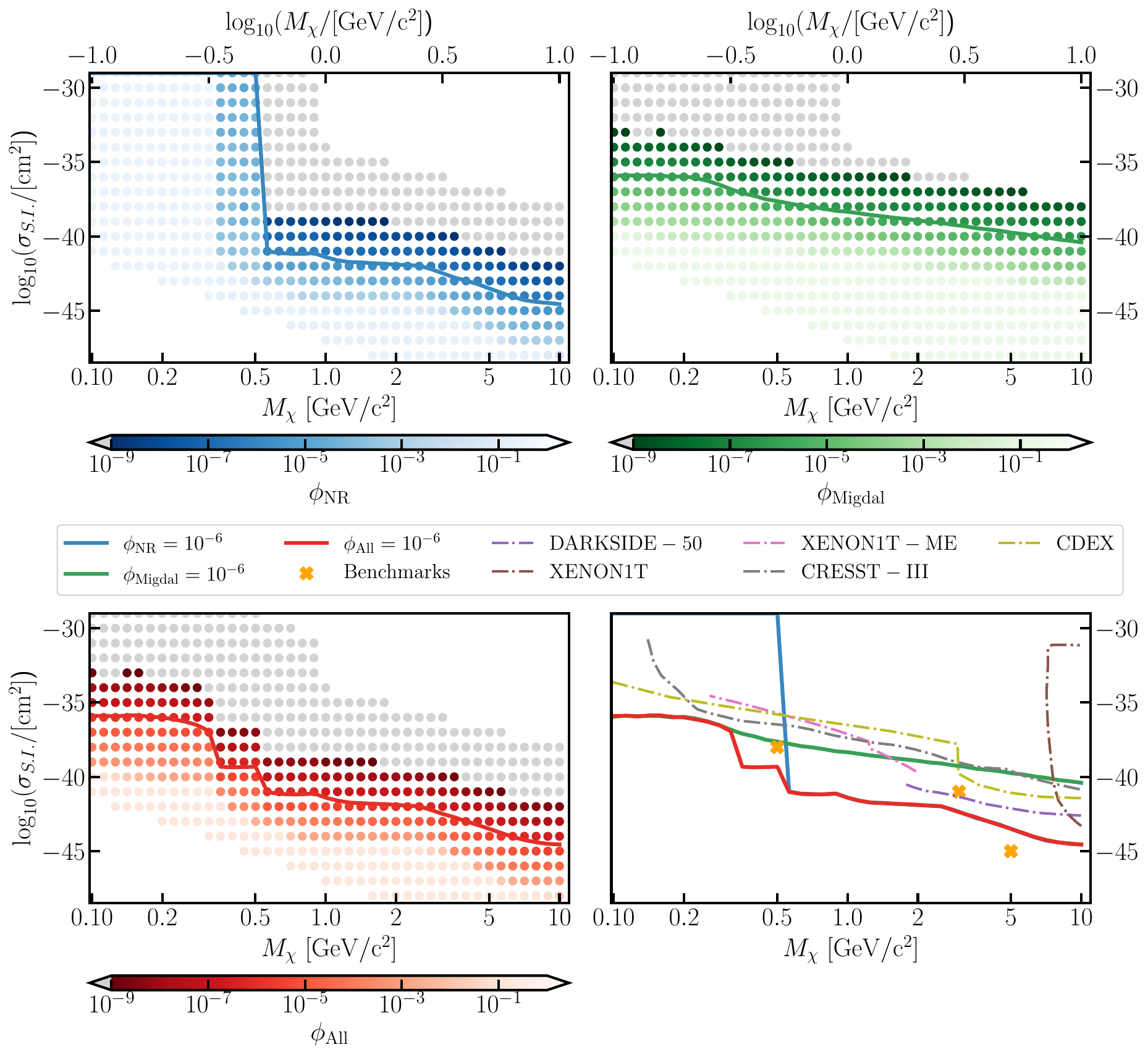}
        \caption{
        Values of $\phi$ for the combined likelihood using the NR (top left), Migdal (top right), or all (bottom right) experiments, where smaller values of $\phi$ indicate a tighter 68 \% CI.
        For each of these results, $\phi$ was interpolated to obtain points where $\phi=10^{-6}$ (solid lines) which are shown again in the comparison panel (bottom right).
        This panel also shows the current experimental exclusion 90\% CL limits of XENON1T Migdal (ME)~\citep{Xe1T_migdal}, XENON1T~\citep{Xe1T_one_ton_year}, CRESST~\citep{Emken_2019}, CDEX~\citep{liu2021studies}, and DarkSide~\citep{darkside50}.
        The benchmark points from \autoref{tab:benchmarks} are plotted as the orange crosses for reference.
        While it is tempting to interpret the lines of $\phi=10^{-6}$ as exclusion limits, this is not correct as elaborated on in the text.
        The results for each of the masses of $\phi_\text{All}$ is interpolated to find the corresponding $\sigmasi$ where $\phi=10^{-6}$ which are the points used in \figref{fig:gridscan_res}.
        Points where $\phi<10^{-9}$ are excluded from the color-scales and all set to gray; these points are all well above the current exclusion limits.
        Points where $\phi\sim\mathcal{O}(10^{-1}-10^{0}$) correspond to Dark Matter parameters that cannot be reconstructed with the 68 \% CI being of similar size as the prior volume.
        \label{fig:gridscan}
        }
    \end{figure*}

    In order to generalize the results as in the sections above, we investigate how the following combined analyses would reconstruct Dark Matter parameters at several WIMP-masses and \cross{}s:
    \begin{itemize}
        \item A combined \textit{NR} analysis using XENONnT-NR and SuperCDMS-NR,
        \item A combined \textit{Migdal} analysis using XENONnT-Migdal and SuperCDMS-Migdal,
        \item A combination of \textit{All} analyses; being    XENONnT-NR, XENONnT-Migdal, SuperCDMS-NR and SuperCDMS-Migdal.
    \end{itemize}
    For each of these analyses, we evaluate $\phi$ for a scan of points in $M_\chi$-$\sigmasi$~space.
    We will refer to these values as $\phi_\text{NR}$, $\phi_\text{Migdal}$, and $\phi_\text{All}$ respectively.
    This allows us to split the contributions of an NR/Migdal analysis to a  fully combined search.

    We perform a grid scan of $M_\chi$ in the range of [0.1,~10]~\gevcsq{} and $\sigmasi$ in the range of [$10^{-47}$,~$10^{-28}$]~\gevcsq{}.
    The points are equally spaced in log space for $\sigmasi$ and $M_\chi$.
    In order to find the parameters resulting in equal $\phi$ for the combination of all detector configurations, the prior range is fixed to [$10^{-2}$,~$10^{2}$]~\gevcsq{} for $M_\chi$ and to [$10^{-53}$,~$10^{-27}$]~\cmsq{} for $\sigmasi$.
    This prior volume is 24\% larger than the priors considered in the previous section (\autoref{tab:benchmarks}), which would therefore yield equally smaller values of $\phi$ for properly reconstructed benchmarks because of the denominator in Eq.~\eqref{eq:eta}.
    Additionally, the number of live points considered here is only 300 in order to save computation time and the values of $\phi$ obtained proved to be similar for 1000 live points.

    \figref{fig:gridscan} shows the results of the grid scan for $M_\chi$ and $\sigmasi$ for the three combinations of analyses.
    Whereas the NR analysis (top left panel) constrains the Dark Matter parameters well for $M_\chi\gtrsim0.5~\gevcsqraw{}$ since $\phi_\text{NR}$ is small, it does not have constraining power below this WIMP-mass.
    The Migdal analyses (top right panel) do have constraining power at these lower WIMP-masses.
    Compared to the NR analysis, the Migdal analysis achieves similar values of $\phi$ above $M_\chi\gtrsim0.5~\gevcsqraw{}$ only at larger $\sigmasi$, meaning that the NR analyses constrain the DM parameters more stringently.

    Generally, for small $M_\chi$ and $\sigmasi$, $\phi\sim\mathcal{O}\left(1\right)$, the combined analyses do not allow constraining the set Dark Matter parameters.
    For large $M_\chi$ and $\sigmasi$, $\phi$ becomes small as the Dark Matter parameters are reconstructed with good precision.\footnote{A significant portion of this parameter space is already excluded by direct detection experiments~\citep{PhysRevD.99.082003, darkside50,Emken_2019,Xe1T_migdal,Xe1T_one_ton_year, liu2021studies}.}

    The combination of all analyses is shown in the bottom left panel, where the contributions of the NR and Migdal analyses are apparent.
    For $M_\chi\gtrsim0.5~\gevcsqraw{}$, the combined result follows the result for NR, while it is dominated by the Migdal result for $M_\chi\lesssim0.3~\gevcsqraw{}$.

    To illustrate this further \figref{fig:gridscan} shows for each of the three combinations the value where $\phi=10^{-6}$.
    While there is nothing particularly special to the value of $\phi=10^{-6}$, it corresponds to values of $\left(M_\chi,~\sigmasi\right)$ that are close to and below the current 90\% confidence level (CL) exclusion limits as illustrated in the bottom right panel of \figref{fig:gridscan}.
    Although it is tempting to interpret the lines where $\phi=10^{-6}$ in this panel as exclusion limits, they are very different.
    Exclusion limits are obtained by doing a one-dimensional fit for a fixed mass and show the (frequentist) 90\% CL upper limit, while  in contrast the lines of $\phi=10^{-6}$ show where a two dimensional fit would be able to reconstruct the WIMP mass and \cross{} simultaneously with good precision.

    To extract points where $\phi=10^{-6}$, we interpolate for each mass in \figref{fig:gridscan} to find the corresponding $\sigmasi$.
    We extract where $\phi=10^{-6}$ in order to obtain $\left(M_\chi,~\sigmasi\right)$-points that are not excluded by experiments at the time of writing \citep{PhysRevD.99.082003,darkside50,Emken_2019,Xe1T_migdal,Xe1T_one_ton_year, liu2021studies}.
    For $\phi_\text{All}$ and $\phi_\text{NR}$ a jump occurs at $M_\chi\sim0.5$~\gevcsq{} as this is near the detection threshold of SuperCDMS-NR; for $\phi_\text{All}$ this is where the transition starts from NR to Migdal being the largest contribution to the total likelihood.

    \begin{figure*}[t!]
        \centering
        \includegraphics[width=0.65\textwidth]{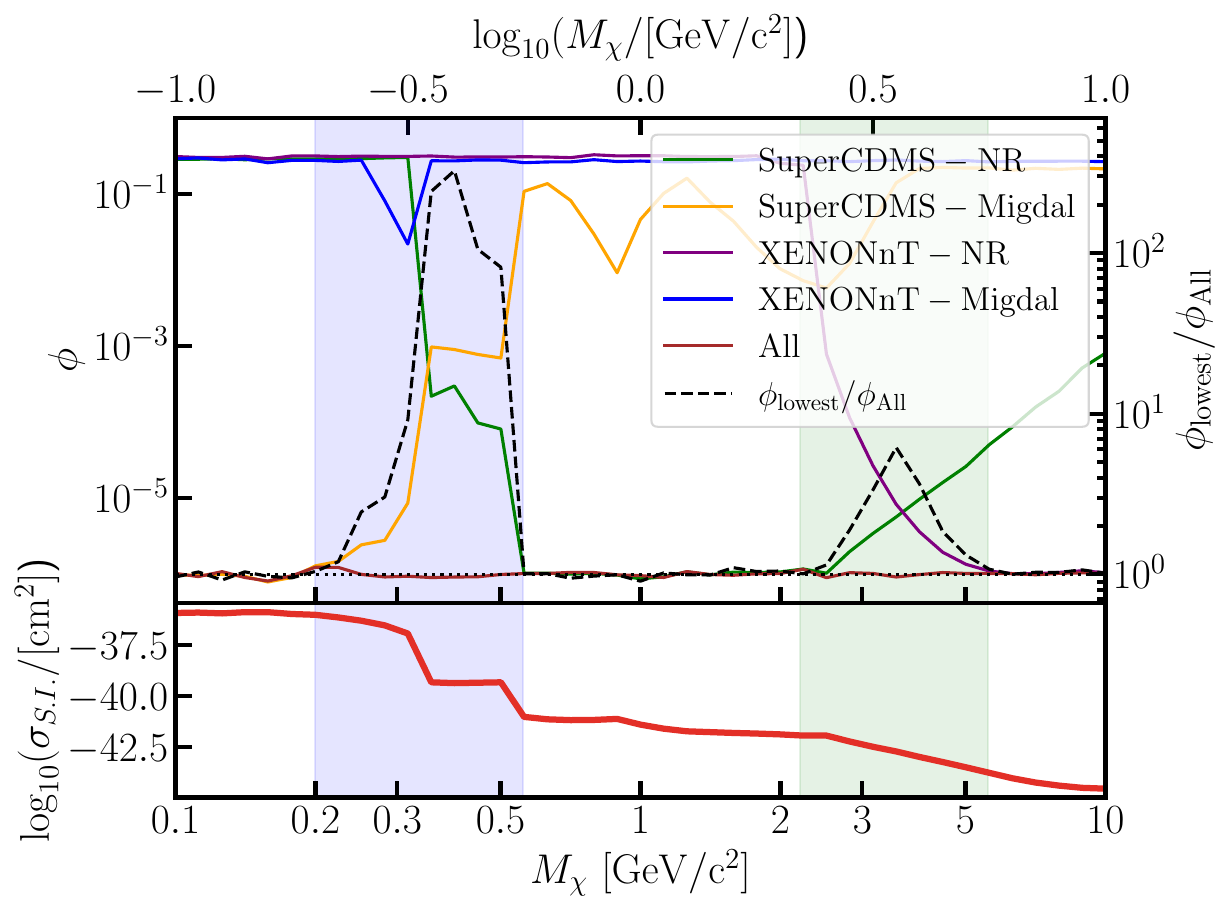}
        \caption{
            Parameter $\phi$ for the four individual detector configurations and $\phi_\text{All}$ (top panel) for the interpolated points from \figref{fig:gridscan}.
            Due to the interpolation, $\phi_\text{All}\sim10^{-6}$ (the horizontal dotted line).
            The right axis (top panel) shows $\phi_\text{lowest}/\phi_\text{All}$, the ratio of the lowest $\phi$ of one of the detector configurations and $\phi_\text{All}$.
            If $\phi_\text{lowest}/\phi_\text{All}\sim1$, the combined likelihood is dominated by the likelihood from one detector configuration as that constrains the parameters well.
            If $\phi_\text{lowest}/\phi_\text{All}\gg1$, this means that the combination of detector configurations is better at constraining the overall likelihood than the individual detector configurations.
            Two mass ranges with high complementarity are shaded and are discussed in the text.
            The bottom panel shows the \cross{} for the masses considered, these correspond to $\phi_\text{All}=10^{-6}$ extracted from the lower left panel of \figref{fig:gridscan}.
            \label{fig:gridscan_res}
        }
    \end{figure*}

    For the $\left(M_\chi,~\sigmasi\right)$-points where $\phi_\text{All}=10^{-6}$, $\phi$ is also calculated for each of the four separate detector configurations to find the detector configuration contributing most to the likelihood.
    If $\phi_\text{All}$ is lower than the $\phi$ of individual detector configurations, this means that the detector configurations are complementary to each other, as in \figref{fig:nr_vs_migdal_3gev_compare}.

    \figref{fig:gridscan_res} evaluates $\phi$ for the individual detector configurations at the points where $\phi_\text{All}=10^{-6}$ in \figref{fig:gridscan}.
    We increase the number of live points back to 1000 from the 300 in considered in \figref{fig:gridscan}.
    Each of the detectors has a mass-range for which it is the most constraining.
    The contribution of XENONnT-NR to the combined likelihood is largest for $M_\chi\gtrsim4~\gevcsqraw{}$ since $\phi_\text{All}\sim\phi_\text{XENONnT-NR}$.
    Similarly, SuperCDMS-NR is most constraining for $M_\chi\sim[0.5,~2.2]~\gevcsqraw{}$, SuperCDMS-Migdal for $M_\chi\lesssim{0.3}~\gevcsqraw{}$.
    We see that the contribution to the combined likelihood from XENONnT-Migdal is small, only achieving values of $\phi_\text{XENONnT-Migdal}\sim\mathcal{O}(10^{-2} - 10^{-1})$ since either XENONnT-NR, SuperCDMS-NR or SuperCDMS-Migdal observes higher rates at the DM parameters considered here.

    At several intermediate masses we find that the combination of detector configurations yields smaller $\phi$ values than the individual detectors.
    For example, between $[2.2,~5.6]~\gevcsqraw{}$, the combination of XENONnT-NR and SuperCDMS-NR yields a smaller value of $\phi$.
    The value of $\phi_\text{All}$ is lower than the individual $\phi$ for the detector configurations of SuperCDMS-NR, SuperCDMS-Migdal and XENONnT-Migdal in the mass range between $\sim[0.2,~0.6]~\gevcsqraw{}$ as all three (mostly SuperCDMS-NR and SuperCDMS-Migdal) are constraining the likelihood.
    In this mass range, a combined analysis will enhance the ability to reconstruct the DM parameters as the $\phi_\text{All}$ is $\mathcal{O}\left(10^1-10^2\right)$ smaller than the smallest $\phi$ for these WIMP masses.

    \section{Conclusion\label{sec:conclusion}}

    We have investigated the potential of two future detectors, XENONnT and SuperCDMS, to discover light WIMP Dark Matter using an NR or Migdal search or combination thereof.
    Using a Bayesian framework to probe the Poisson likelihood, the posterior distributions of benchmark points were obtained for WIMP masses of $5,~3\text{ and }0.5~\gevcsqraw{}$ and \cross{} of $10^{-45},~10^{-41}\text{ and }10^{-38}$~\cmsq{} respectively.
    For $5$~\gevcsq{} (\autoref{fig:nr_vs_migdal_5gev}), XENONnT-NR constrained the Dark Matter parameters most, whereas for $0.5$~\gevcsq{} (\autoref{fig:nr_vs_migdal_05gev}) this was done by SuperCDMS-Migdal.
    At an intermediate mass of $3$~\gevcsq{} (\autoref{fig:nr_vs_migdal_3gev}) the parameter $\phi$ reduces for the posterior of the combined likelihood by a factor of 54 for XENONnT-NR and 2.1 for SuperCDMS-NR  (\autoref{fig:nr_vs_migdal_3gev_compare}).

    More generally, we probed a large parameter space in $\left(M_\chi,~\sigmasi\right)$ to find the set of DM parameters where a combined inference of the NR, Migdal, all combined-analyses would be able to reconstruct those DM parameters to an equally sized 68\% CI (\figref{fig:gridscan}).
    Using those points, we observed several regions in which one of the detection configurations was outperforming the other detector configurations (\figref{fig:gridscan_res}).
    Near the detection threshold of XENONnT-NR ($\sim[2.2,~5.6]~\gevcsqraw{}$), the combination with SuperCDMS-NR helps in reconstructing the DM parameters.
    The largest complementarity can be found for SuperCDMS-NR,  SuperCDMS-Migdal, and to a lesser extent, XENONnT-Migdal in the mass range between $\sim[0.2,~0.6]~\gevcsqraw{}$.

    In future work, several effects may be worth exploring.
    One of the most important parameters for XENONnT is the energy threshold.
    Experiments are cautious with claiming discoveries near detection thresholds as threshold effects are difficult to model fully.
    An interesting study would be to take the value of the energy threshold into account as a nuisance parameter in Eq.~\eqref{eq:parameters_theta}.
    Similarly, as was done previously in Ref.~\citep{Pato_complementarity}, it is worth doing the same for the astrophysical DM parameters.
    While this has been well-studied for NR searches, their effect on Migdal searches have not been investigated.
    Finally, the Earth shielding effect \citep{verne_paper} should be taken into account when discussing the ability to detect strongly interacting Dark Matter, either at the very small or very large WIMP-masses where large \cross{}s are not excluded by experimental results.

    We have demonstrated the complementarity of two planned Dark Matter direct detection experiments to observe light Dark Matter through a combination of Migdal and standard NR searches.
    These results highlight in particular that over certain WIMP mass ranges the combination of standard NR and Migdal searches can lead to tighter constraints on the Dark Matter parameters than from either analysis alone.


    \acknowledgments

    B.J.K.\ thanks the Spanish Agencia Estatal de Investigaci\'on (AEI, Ministerio de Ciencia, Innovación y Universidades) for the support to the Unidad de Excelencia Mar\'ia de Maeztu Instituto de F\'isica de Cantabria, ref. MDM-2017-0765. We gratefully acknowledge support from the Dutch Research Council (NWO).

    \appendix

    \section{Energy scales}

    In this appendix we review several details required for converting the energy scales relevant for the detectors in this work.

    \subsection{Lindhard quenching\label{ap:lindhard}}

    The two detectors of interest (SuperCDMS-SNOLAB and XENONnT) both use ionization signals caused by interactions to characterize the type of interaction (ER or NR) within the target volume.
    In xenon, germanium and silicon, an ER of a given energy will result in more detectable ionization energy than an NR of the same energy due to nuclear quenching \citep{lindhard1963integral,akerib2016low}.
    We adopt the following notation for the ER recoil energy $\Eer$ and the NR recoil energy $\Enr$.
    In order to compare NR and ER energies it is often useful to calculate how much ionization energy a nuclear recoil would have deposited if the recoil was an electronic recoil: the electronic equivalent energy ($E_\text{ee}$).
    Using the Lindhard factor $L$ \citep{lindhard1963integral,akerib2016low},

    \begin{align}
        L(\Enr) &= \frac{k\ g(\epsilon)}{1+k g(\epsilon)},\label{eq:lindhard}\\
        g(\epsilon)&=3\epsilon^{0.15}+0.7\epsilon^{0.6}+\epsilon,\nonumber\\
        \epsilon &= 11.5\frac{\Enr}{\text{keVnr}}\,Z^{-7/3}\nonumber\,,
    \end{align}
    we can convert $\Enr$ to $E_\mathrm{ee}$:
    \begin{equation}
        E_\text{ee} = L(\Enr) \cdot \Enr\,. \label{eq:ee_to_nr}
    \end{equation}
    Here, $k$ is a detector specific parameter and $Z$ the atomic number of the target material. From Eq.~\eqref{eq:lindhard}, we can directly see that $L<1$.
    The Lindhard factor is used to convert $\Enr$ into $E_\text{ee}$ and vice versa in the methods section (\autoref{sec:method}).
    
    Following~\citep{quenching_effect_si_ge}, we rewrite Eq.~\eqref{eq:lindhard} to take the atomic binding energy into account for semiconductor materials:
    \begin{align}
        L(\Enr) &= \frac{k\ g(\epsilon')}{1+k g(\epsilon')}=\frac{\epsilon' - \bar{\nu}(\epsilon')}{\epsilon'}\label{eq:lindhard_sige}\,,\\
        \bar{v}(\epsilon')&=\bar{v}_L + C_0{\epsilon'}^\frac{1}{2}+C_1+u\nonumber\,,\\
        \bar{v}_L(\epsilon')&=\frac{\epsilon'}{1+kg(\epsilon')}\nonumber\,,\\
        u&= 11.5\frac{\Enr}{\text{keVnr}}\,Z^{-7/3}U\nonumber\,,\\
        \epsilon' &= \epsilon - u\nonumber\,,
    \end{align}
    where $U$ is the energy lost to disruption of atomic bonding, $C_0$ and $C_1$ are material specific parameters. For $C_0=C_1=0$ and $U=0$~\kev, Eq.~\eqref{eq:lindhard_sige} reduces to Eq.~\eqref{eq:lindhard}. We use the best fit parameters as obtained in Ref.~\cite{quenching_effect_si_ge}. For Si we take $C_0=9.1$\scinot{-3}, $C_1=3.3$\scinot{-5} and $U=0.15$~\kev. For Ge, we take $C_0=3.0$\scinot{-4}, $C_1=6.2$\scinot{-6} and $U=0.02$~\kev. We assume a value of $k$ of $0.162$ for Ge and $0.161$ for Si \citep{quenching_effect_si_ge} in Eq.~\eqref{eq:lindhard_sige}.

    \subsection{SuperCDMS energy-resolution and~-threshold\label{ap:e_supercdms}}

    In this appendix, the two relevant energy scales for SuperCDMS are discussed as well as how the values for \autoref{tab:det_params} for the energy-resolution and -threshold are obtained.

    There are two energy scales in the SuperCDMS experiment that relate to the ER/NR recoil energy scales \citep{superCDMSsnolab}, namely the phonon energy $E_{ph}$ and the ionization energy $E_Q$, where the latter is given by:\footnote{Here, we are only considering ``bulk events" that have a correction factor $\eta=1$ in Equations 3 and 4 of Ref.~\citep{superCDMSsnolab}.}
    \begin{equation}
        E_{Q,\, \text{nr}} = y(\Enr)\cdot \Enr\,, \label{eq:eq_nr}
    \end{equation}
    where $y(\Enr)$ is the ionization yield, which is set to be equal to $L(\Enr)$ for large enough $\Enr$.
    For ERs, where $y=1$, we can explicitly rewrite this as:
    \begin{equation}
        E_{Q,\, \text{er}} = \Eer\,. \label{eq:eq_er}
    \end{equation}


    Additionally, the phonon energy scale is given by:
    \begin{align}
        E_\text{ph, nr} &= \Enr + E_\text{Luke, nr}\nonumber \\
        &=\Enr \left(1+\frac{y(\Enr)e\Delta V}{\delta}\right)\,,\label{eq:eph_nr}\\
        E_\text{ph, er} &= \Eer + E_\text{Luke, er} \nonumber\\
        &=\Eer \left(1+ \frac{e\Delta V}{\delta}\right)\,,\label{eq:eph_er}
    \end{align}
    where the $ E_\text{Luke}$-term is the signal generated through the Luke-Neganov effect \citep{superCDMSsnolab}, $\delta$ is the average energy required to make an electron-hole pair ($3.0~\ev$ for Ge and $3.82~\ev$ for Si) and $e\Delta V$ is the work done to move one charge through a crystal, which depends on the bias voltage applied to the detector.
    The value of  $e\Delta V/\delta$ depends on the detector design and is 1.6 (Ge) or 2.7 (Si) for IZIP, and 26 (Ge) or 33 (Si) for HV.
    As such a relatively modest $\Eer$ can correspond to a large $E_\text{ph}$.

    For Migdal, the recoil spectrum is computed in $\Eer$.
    However, in Ref.~\citep{superCDMSsnolab}, the resolution and energy thresholds are given in $E_{ph}$.
    We need to convert the energy threshold by inverting Eq.~\eqref{eq:eph_er} and substituting the $E_{ph}$ (from Table VIII in Ref. \citep{superCDMSsnolab}).

    Similar to the energy threshold, the energy resolution is given in the phonon resolution $\sigma_\text{ph}$.
    This resolution is in the order $5-50~\ev$. We relate the phonon resolution $\sigma_{ph}$ to the ER resolution $\sigma_\text{er}$ using Eq.~\eqref{eq:eph_er}:
    \begin{equation}
        \sigma_\text{er}=\sigma_\text{ph}/\left(1+\frac{e\Delta V}{\delta}\right)\,.
    \end{equation}

    For the NR search in SuperCDMS we need to have the relevant energy resolutions and thresholds
    By inverting Eq.~\eqref{eq:eph_nr}, we can obtain the values listed for the NR energy threshold in Ref. \citep{superCDMSsnolab}, which are directly used in \autoref{tab:det_params}.
    For the NR case, we need to distinguish between the ionization resolution relevant for the iZIP detectors and the phonon resolution, relevant for the HV detectors.
    As such, if we treat $\sigma_\text{ph,nr}$ as the uncertainty on $E_\text{ph,nr}$, we can propagate the resolution $\sigma_\text{ph,nr}$ to $\sigma_\text{nr}$ as:
    \begin{equation}
        \sigma_\text{nr}=\frac{\text{d} \Enr}{\text{d}E_\text{ph,nr}}\sigma_\text{ph, nr}\label{eq:sigma_ph_nr}
        \,,
    \end{equation}
    and resolution of $\sigma_{Q,\,\text{nr}}$ to $\sigma_\text{nr}$ as:
    \begin{equation}
        \sigma_\text{nr}=\frac{\text{d} \Enr}{\text{d}E_{Q,\,\text{nr}}}\sigma_{Q,\,\text{nr}}\,,\label{eq:sigma_q_nr}
    \end{equation}
    where Eq.~\eqref{eq:sigma_ph_nr} applies to the HV detectors and Eq.~\eqref{eq:sigma_q_nr} to the iZIP detectors.
    We solve Eqs. (\ref{eq:sigma_ph_nr}-\ref{eq:sigma_q_nr}) numerically.
    From Eqs. (\ref{eq:sigma_ph_nr}-\ref{eq:sigma_q_nr}), we see that the energy resolution $\sigma_\text{nr}$ has an energy dependence through the ionization yield $y(\Enr)$ even though $\sigma_\text{ph, nr}$ and $\sigma_\text{Q, nr}$ are assumed to be energy independent.


    \bibliographystyle{JHEP}
    \bibliography{main.bib}

\end{document}